\documentclass[a4paper,11pt]{article}

\usepackage{amssymb,amsbsy,amsmath,amsthm}

\newcommand{\Com}{\mathop{\mathbb C}\nolimits}            % doubled C
\newcommand{\Real}{\mathop{\mathbb R}\nolimits}           % doubled R
\newcommand{\posReal}{\mathop{\mathbb R}^+\nolimits}      % doubled R+
\newcommand{\Nat}{\mathop{\mathbb N}\nolimits}            % doubled N
\newcommand{\proj}{\mathop{\mathcal{P}_1^\bot}\nolimits}  % projector

\newcommand{\ie}{\emph{i.e.}}                             % i.e.
\newcommand{\eg}{\emph{e.g.}}                             % e.g.
\newcommand{\cf}{\emph{cf.}}                              % cf.
\newcommand{\sgn}{\mathop{\mathrm{sgn}}\nolimits}         % sgn
\newcommand{\tr}{\mathop{\mathrm{tr}}\nolimits}           % tr
\newcommand{\Dom}{\mathop{\mathrm{Dom}}\nolimits}         % Dom

\newcommand{\layer}{\mathop{\mathcal{L}}\nolimits}        % layer mapping
\newcommand{\trans}{\mathop{I}\nolimits}                  % (-a,a)
\newcommand{\TotK}{\mathop{\mathcal{K}}\nolimits}         % TotK
\newcommand{\Euler}{\mathop{\gamma_\mathrm{E}}\nolimits}  % Euler's gamma

\newcommand{\si}{\mathop{L^1}\nolimits}                   % L1-space
\newcommand{\sii}{\mathop{L^2}\nolimits}                  % L2-space
\newcommand{\s}{\mathop{L}\nolimits}                      % L-space
\newcommand{\sinf}{\mathop{L^\infty}\nolimits}            % Linf-space
\newcommand{\sobi}{\mathop{W_0^{1,2}}\nolimits}           % Sobolov space 1,2,0
\newcommand{\Comp}{\mathop{C_0^\infty}\nolimits}          % C_0^\infty
\newcommand{\Smooth}{C}                                   % C

\newcommand{\Assi}{\mbox{$\langle\textsf{a1}\rangle$}}
\newcommand{\Assii}{\mbox{$\langle\textsf{a2}\rangle$}}
\newcommand{\Assiii}{\mbox{$\langle\textsf{a3}\rangle$}}
\newcommand{\AssV}{\mbox{$\langle\textsf{a0}\rangle$}}
\newcommand{\AssA}{\mbox{$\langle\textsf{a1--3}\rangle$}}

\newcommand{\AssDi}{\mbox{$\langle\textsf{d1}\rangle$}}
\newcommand{\AssDii}{\mbox{$\langle\textsf{d2}\rangle$}}
\newcommand{\AssDiii}{\mbox{$\langle\textsf{d3}\rangle$}}
\newcommand{\AssDiv}{\mbox{$\langle\textsf{d4}\rangle$}}
\newcommand{\AssRiy}{\mbox{$\langle\textsf{r1,2}\rangle$}}

\newcommand{\AssRiii}{\mbox{$\langle\textsf{r3}\rangle$}}
\newcommand{\AssD}{\mbox{$\langle\textsf{d1--4}\rangle$}}
\newcommand{\AssR}{\mbox{$\langle\textsf{r1--3}\rangle$}}

\def\OMIT#1{}

\newtheorem{claim}{Claim}[section]
\newtheorem{prop}[claim]{Proposition}                     % Proposition
\newtheorem{thm}[claim]{Theorem}                          % Theorem
\newtheorem{lemma}[claim]{Lemma}                          % Lemma
\newenvironment{PROOF}
  {\begin{proof}[\textsc{Proof:}]}{\end{proof}}           % PROOF

\theoremstyle{definition}
\newtheorem*{rem}{Remark}                                 % Remark
\newtheorem*{rems}{Remarks}                               % Remarks

\theoremstyle{remark}
\begin{document}

\title{\textbf{Bound States in Mildly Curved Layers}}
\author{P.~Exner$^{a,b}$
    and D.~Krej\v{c}i\v{r}\'{\i}k$^{a,c-e}$}
\date{March 15, 2001}
\maketitle
\vspace{-5.5ex}
{\small\em
\begin{description}
\item[$a)$] Nuclear Physics Institute, Academy of Sciences,
            25068 \v{R}e\v{z} near Prague
\vspace{-1.2ex}
\item[$b)$] Doppler Institute, Czech Technical University,
            B\v{r}ehov{\'a}~7, 11519 Prague
\vspace{-1.2ex}
\item[$c)$] Faculty of Mathematics and Physics, Charles University,
            V~Ho\-le\-\v{s}o\-vi\v{c}\-k\'ach~2, 18000 Prague, Czech Republic
\vspace{-1.2ex}
\item[$d)$] Centre de Physique Th\'eorique, CNRS,
            13288 Marseille-Luminy
\vspace{-1.2ex}
\item[$e)$] PHYMAT, Universit\'e de Toulon et du Var,
            83957 La Garde, France
\vspace{-1.2ex}
\end{description}
}
\begin{center}
exner@ujf.cas.cz, krejcirik@ujf.cas.cz
\end{center}
\smallskip
\begin{abstract}
\noindent It has been shown recently that a nonrelativistic
quantum particle constrained to a hard-wall layer of constant
width built over a geodesically complete
simply connected noncompact curved surface
can have bound states provided the surface is not a plane. In this
paper we study the weak-coupling asymptotics of these bound
states, \ie, the situation when the surface is a mildly curved
plane. Under suitable assumptions about regularity and decay of
surface curvatures we derive the leading order in the ground-state
eigenvalue expansion. The argument is based on Birman-Schwinger
analysis of Schr\"odinger operators in a planar hard-wall layer.
%
%\medskip\noindent
%\textbf{Key-Words:} waveguides, layers, Dirichlet Laplacian, bound
%states, weak coupling, Birman-Schwinger analysis, Schr\"odinger
%operators, curvature, mean curvature, mildly curved surfaces,
%asymptotically planar character
%
\end{abstract}

\setcounter{equation}{0}
%%%%%%%%%%%%%%%%%%%%%%
\section{Introduction}
%%%%%%%%%%%%%%%%%%%%%%
The investigation of quantum particles constrained to a spatial
region $\Omega$ of a prescribed shape, in particular, relations
between its spectral properties and the geometry of $\Omega$
became an attractive problem when the technology progress made it
possible to fabricate various mesoscopic systems for which this is
a reasonable model -- see, \eg, \cite{LCM}. Curvature induced bound
states in hard-wall strips and tubes have been demonstrated more
than a decade ago~\cite{ES1} and studied subsequently in numerous papers
-- \cf~\cite{DE1,LCM} and references therein.

Much less attention was paid to quantum mechanics in layers, apart
of the trivial planar case. While for an experimenter it is easier
to prepare a semiconductor film on a curved substrate than to
fabricate a quantum wire, from the mathematical point of view the
opposite is true, and solving the Schr\"odinger equation in a
nontrivial layer is more complicated than the corresponding
problem in a tube. It was noticed long time ago that in the formal
limit of zero width the layer curvatures give rise to an effective
attractive potential~\cite{KJ}
but the first results for curved finite-width layers, with the
particle Hamiltonian being a multiple of the Dirichlet
Laplacian~$-\Delta_D^\Omega$, appeared only recently.

We restrict ourselves to the case when $\Omega$ is non-compact and
nontrivially curved. The first proof of existence of geometrically
induced bound states was given in~\cite{DEK} under the assumption
that the planar layer is curved only locally. A more general case
of curved layers which are \emph{asymptotically} planar in the
sense that the curvatures of the generating surface vanish at
large distances has been discussed in~\cite{DEK1}. We have derived
there several sufficient conditions for the existence of bound
states expressed in terms of geometric quantities characterizing
the reference surface. While these conditions cover a wide class
of layers, they do not represent an ultimate result: it is not
clear, \eg, whether bound states exist in layers built over
surfaces of positive total Gauss curvature unless the latter are
thin enough or endowed with a cylindrical symmetry.

After demonstrating their existence one is naturally interested in
properties of the curvature-induced bound states. A particular
question which we are going to address in the present paper
concerns the weak-coupling regime. Since in our case the binding
comes from the curvature alone, the described situation is
expected to occur in \emph{mildly curved} layers. We will show
that the layer has then a unique eigenvalue and derive an
asymptotic expansion for the gap between this eigenvalue and the
threshold of the essential spectrum. The leading term in this
formula will depend on the mean curvature of the reference
surface.

Let us describe briefly the contents of the paper. In the next
section we will give a precise formulation of the problem and
describe the main result summarized in Theorem~\ref{Thm.WCL}. The
rest is devoted to the proof. The strategy is adopted from the
``lower-dimensional'' case of curved strips and tubes analyzed
in~\cite[Sec.~4]{DE1}. First, in Section~\ref{Sec.WCL}, we
consider Schr\"odinger operators acting in a planar layer built
over~$\Real^2$. We derive a necessary and sufficient condition
under which such an operator has a bound state in the weak
coupling limit and find the asymptotic expansion of the
eigenvalue. The result contained in Theorem~\ref{Thm.Expansion} is
of an independent interest; we prove it in a greater generality
for a layer in $\Real^{2+m},\, m\ge 1$. Next, in
Section~\ref{Sec.BS}, we apply it to the case of mildly curved
quantum layers. We express the operator~$-\Delta_D^\Omega$ in the
coordinates~$(x,u)$, where~$x\equiv(x^1,x^2)\in\Real^2$
and~$u\in\trans:=(-a,a)$ with~$a>0$ parametrize the surface and
its normal space, respectively, and pass to a unitarily equivalent
operator with an effective potential, which can be consecutively
estimated by operators to which Theorem~\ref{Thm.Expansion} can be
applied . We will also show how the leading term in the expansion
looks like in the case of a thin layer.

\setcounter{equation}{0}
%%%%%%%%%%%%%%%%%%%%%%
\section{The results} \label{res}
%%%%%%%%%%%%%%%%%%%%%%
To give a precise statement of the main result we need first to
specify what we mean by mildly curved quantum layers. Let a family
of surfaces $\Sigma_\varepsilon:=p\,(\Real^2)$ be given by a Monge
patch
\begin{equation}\label{surface}
  p:\Real^2\to\Real^3, \qquad
  p\,(x^1,x^2;\varepsilon)
  :=\left(x^1,x^2,\varepsilon f(x^1,x^2)\right),
\end{equation}
where~$f$ is supposed to be a $\Smooth^4$-smooth function. Here
$\varepsilon>0$ is the parameter which controls the deformation;
it is supposed to be small so that~$\Sigma_\varepsilon$ is a
mildly curved plane. The cross-product of the tangent vectors
$p_{,\mu}:=\partial p/\partial x^\mu$, $\mu=1,2$, defines a unit
normal field~$n$ on~$\Sigma_\varepsilon$. We
put~$\Omega_0:=\Real^2\times\trans$ and define a layer
$\Omega_\varepsilon:=\layer(\Omega_0)$ of width~$d=2 a$ over the
surface~$\Sigma_\varepsilon$ by virtue of the mapping
$\layer:\Omega_0\to\Real^3$ which acts as
\begin{equation}\label{layer}
  \layer(x,u;\varepsilon)
  :=p(x;\varepsilon)+u\,n(x;\varepsilon).
\end{equation}
The layer in question is thus the spatial region closed between
two parallel surfaces -- \cf~\cite[Prob.~12 of Chap.~3]{Spivak3}
-- represented by~$p \pm a n$. In this sense, we will hereafter
refer to the surface coordinates~$(x^1,x^2)$ as to longitudinal
variables, while~$u$ will be the transverse variable.

We consider a nonrelativistic spinless particle confined
to~$\Omega_\varepsilon$ which is free within it, and suppose that
the boundary of the layer is a hard wall, \ie, the wavefunctions
satisfy the Dirichlet boundary condition there. For the sake of
simplicity we set Planck's constant $\hbar=1$ and the mass of the
particle $m=\frac{1}{2}$. Then the Hamiltonian of the system can
be identified with the Dirichlet Laplacian
$-\Delta_D^{\Omega_\varepsilon}$ on~$\sii(\Omega_\varepsilon)$,
which is defined for an open
set~$\Omega_\varepsilon\subset\Real^3$ as the Friedrichs extension
of the Laplacian acting on~$\Comp(\Omega_\varepsilon)$ --
\cf~\cite[Sec.~XIII.15]{RS4} or~\cite[Chap.~6]{Davies}. The domain
of the closure of the corresponding quadratic form is the Sobolev
space~$\sobi(\Omega_\varepsilon)$.

Next we must introduce some notation and formulate assumptions
about the function~$f$, in addition to the smoothness requirement
mentioned above. We consider asymptotically planar surfaces --
\cf~\cite{DEK1} -- which means that the Gauss~$K$ and mean~$M$
curvatures of~$\Sigma_\varepsilon$ vanish at large distances from
a fixed point. This will hold if we require
\begin{description}
\item \framebox{\AssDi\ $f_{,\mu}\in\sinf(\Real^2)$}
      \quad together with \quad
      \framebox{\AssDii\ $f_{,\mu\nu}\to 0$ \ as $|x|\to\infty$}
\end{description}
for~$\mu,\nu\in\{1,2\}$. This allows us to localize the essential
spectrum. Adapting from~\cite[Thm.~4.1]{DEK1} a simple argument
based on a Neumann bracketing one finds that
$\inf\sigma_\mathrm{ess}(-\Delta_D^{\Omega_\varepsilon})
\geq\kappa_1^2$. Here~$\{\kappa_j^2:=(\kappa_1
j)^2\}_{j=1}^\infty$, with~$\kappa_1:=\pi/d$, are the eigenvalues
of~$-\Delta_D^{\trans}$; the corresponding eigenfunctions will be
denoted by~$\chi_j$. Assuming in addition
\begin{description}
\item \framebox{\AssDiii\ $f_{,\mu\nu\!\rho}\to 0$ \ as $|x|\to\infty$}
      \quad and \quad
      \framebox{\AssDiv\ $f_{,\mu\nu\!\rho\sigma}\in\sinf(\Real^2)$}
\end{description}
for~$\mu,\nu,\rho,\sigma\in\{1,2\}$ we will be able to prove in
Section~\ref{Sec.BoundState} that the bound is sharp,
\begin{equation}
  \inf\sigma_\mathrm{ess}(-\Delta_D^{\Omega_\varepsilon})
  =\kappa_1^2.
\end{equation}
We have also to require that the derivatives of~$f$ satisfy the
following integrability hypotheses:
\begin{description}
\item \framebox{
      \AssRiy\ $f_{,\mu\nu}\,$, $\:f_{,\mu\nu\!\rho}
      \,\in\sii\left(\Real^2,(1+|x|^\delta)\,dx\right)$,
      $\:$respectively
      }
\end{description}
and
\begin{description}
\item \framebox{
      \AssRiii\ $f_{,\mu\nu\!\rho\sigma}$
      $\in\si\left(\Real^2,(1+|x|^\delta)\,dx\right)$
      } \qquad
      for some~$\delta>0$
\end{description}
and~$\mu,\nu,\rho,\sigma\in\{1,2\}$. The main result of the
present work reads as follows:
\begin{thm}\label{Thm.WCL}
Let~$\Omega_\varepsilon$ be a family of layers generated by the
surfaces~$\Sigma_\varepsilon$  given by~\emph{(\ref{surface})}.
Suppose that the function~$f\in\Smooth^4(\Real^2)$ satisfies the
hypotheses~\emph{\AssD} and~\emph{\AssR}. If~$\Sigma_1$ is not
planar, then for all~$\varepsilon$ small enough
$-\Delta_D^{\Omega_\varepsilon}$ has exactly one isolated
eigenvalue~$E(\varepsilon)$ below the essential spectrum.
Moreover, it can be expressed as
$$
  E(\varepsilon)=\kappa_1^2-e^{2 w(\varepsilon)^{-1}},
$$
where~$w(\varepsilon)$ has the following asymptotic expansion
$$
  w(\varepsilon)=-\varepsilon^2
  \sum_{j=2}^\infty (\chi_1,u\chi_j)_{\!\trans}^2\,
  \left(\kappa_j^2-\kappa_1^2\right)^2
  \int_{\Real^2} \frac{|\widehat{m_0}(\omega)|^2}
  {|\omega|^2+\kappa_j^2-\kappa_1^2}\,d\omega
  +\mathcal{O}(\varepsilon^{2+\gamma})
$$
with~$\gamma:=\min\{1,\delta/2\}$. Here~$m_0$ is the lowest-order
term in the expansion of the mean curvature
of~$\Sigma_\varepsilon$ w.r.t.~$\varepsilon$ -- \cf~\emph{(\ref{km})}.
\end{thm}
\begin{rems}
(a) The subscript~``$\trans$'' indicates the norm and the inner
product in the space~$\sii(\trans)$. The sum runs in fact over
even~$n$ only because one integrates over the interval~$I=(-a,a)$
on which the function $u\mapsto \chi_1(u) u \chi_j(u)$ is odd for
odd~$j$.

(b) The expression for the leading-term coefficient~$w_1$ in the
expansion $w(\varepsilon)=:\varepsilon^2 w_1
+\mathcal{O}(\varepsilon^{2+\gamma})$\ does not have a very
transparent structure, however, for thin layers it can be
rewritten as
\begin{equation}\label{ThinExpansion}
  w_1=-\frac{1}{2\pi}\,\|m_0\|^2
  +\frac{\pi^2-6}{24\pi^3}\,\|\nabla m_0\|^2 d^2
  +\mathcal{O}(d^4).
\end{equation}
This formula is instructive because the first term comes from the
surface attractive potential~$K-M^2$ which dominates the picture
in this case, while the~$\mathcal{O}(d^4)$ error term expresses
the contribution of higher transverse modes. We refer to
Section~\ref{Sec.Thin} for more details.
\end{rems}
%

%%%%%%%%%%%%%%%%%%%%%%%%%%%%%%%%%%%%%%%%%%%%%%%%%%
\begin{sloppy}
\section{Weakly Coupled Schr\"odinger Operators in
     a Planar Layer}\label{Sec.WCL}
\end{sloppy}\noindent
%%%%%%%%%%%%%%%%%%%%%%%%%%%%%%%%%%%%%%%%%%%%%%%%%%
Let~$M$ be an open connected precompact set in $\Real^m$, $m\geq
1$. The object of our interest in this section will be the
operator
\begin{equation}\label{WCLHamiltonian}
  H_\lambda=-\Delta_D+\lambda V
  \qquad\textrm{with}\quad\lambda>0 \
  \quad\ \textrm{on}\quad
  \mathcal{H}:=\sii(\Real^2)\otimes\sii(M),
\end{equation}
where~$-\Delta_D$ is the Dirichlet Laplacian on~$\Real^2\times M$
defined as the closure of $-\Delta\otimes I_m+I_2\otimes
-\Delta_D^M$. In the last expression the unindexed~$-\Delta$ stays
for the Laplace operator in~$\Real^2$. The potential~$V$ is
supposed to be $H_0^\frac{1}{2}$-bounded, in other words
\vspace{0.8ex}
\begin{description}
\item\framebox{\Assi\
  $\exists\, a,b \geq 0 \quad \forall\psi\in\Dom Q_0: \quad
  \|V\psi\| \leq a \|\psi\| + b\, \big\|H_0^\frac{1}{2}\psi\big\|\:$}
\end{description}
where~$\Dom Q_0=\sobi(\Real^2\times M)$ is the form domain of~$H_\lambda$.

The Dirichlet Laplacian~$-\Delta_D^M$ on~$\sii(M)$ has a purely
discrete spectrum consisting of eigenvalues
$\kappa_1^2<\kappa_2^2\le\dots\le\kappa_j^2<\dots$; the
corresponding normalized eigenfunctions will be denoted
as~$\chi_j$, where~$j=1,2,\dots\:$.
The lowest eigenvalue is, of course, simple and the eigenfunction~$\chi_1$
can be chosen positive -- \cf~\cite[Sec.~XIII.~12]{RS4}.
Important for our purpose are the transverse projections of the
potential,
$$
  V_{jj'}:=\int_M \bar{\chi}_j(y) \,
  V(\cdot,y) \, \chi_{j'}(y) \, dy
$$
and the analogous quantities for other functions on~$\Real^2\times
M$, in particular for~$|V|$. We shall adopt the assumptions
\begin{description}
\item\framebox{\Assii\ $|V|_{11}\in\s^{1+\delta}\left(\Real^2\right)$}
\quad\textrm{and}\quad \framebox{\Assiii\
$|V|_{11}\in\si\left(\Real^2,(1+|x|^\delta)\,dx\right)$}
\end{description}
for some~$\delta>0$. We also suppose that the essential spectrum
of~$H_\lambda$ does not start below the lowest transverse-mode
threshold, \ie,
\begin{description}
\item\framebox{\AssV\ $\inf\sigma_\mathrm{ess}(H_\lambda) \geq \kappa_1^2$}
\end{description}
This is true, in particular, if~$V$ vanishes at large distances
from a fixed point.

The goal of this section is to show that for~$\lambda$ small
enough the discrete spectrum of~$H_\lambda$ below~$\kappa_1^2$ is
not empty provided the projection of~$V$ onto~$\chi_1$ is not
repulsive in the mean. This part of spectrum then contains only
one eigenvalue~$E(\lambda)$ for such a small~$\lambda$ and it
approaches~$\kappa_1^2$ as~$\lambda$ tends to zero. The last claim
follows from the following elementary fact.
\begin{prop}\label{Proposition}
Assume~\emph{\Assi}. Then there are positive constants~$\lambda_0$
and~$c$ such that for all~$\lambda\in(0,\lambda_0)$ we have
$H_\lambda\geq\kappa_1^2-c\lambda$.
\end{prop}
\begin{PROOF}
Using the Schwarz inequality and the assumption~\Assi\/
together with the inequality between the geometric and arithmetic means,
we have for all~$\psi\in\Dom Q_0$ the bound
\begin{eqnarray*}
  Q_\lambda[\psi]:=\big\|H_\lambda^\frac{1}{2}\psi\big\|^2
&\geq& \big\|H_0^\frac{1}{2}\psi\big\|^2-\lambda\|\psi\|\,\|V\psi\| \\
&\geq& (1-\lambda\,b/2)\,\big\|H_0^\frac{1}{2}\psi\big\|^2
  -\lambda\,(a+b/2)\,\|\psi\|^2.
\end{eqnarray*}
Since~$\big\|H_0^\frac{1}{2}\psi\big\|\geq\kappa_1\,\|\psi\|$, we obtain
the assertion by putting~$\lambda_0:=2/b$
and~$c:=a+(1+\kappa_1^2)\,b/2$.
\end{PROOF}
%

%%%%%%%%%%%%%%%%%%%%%%%%%%%%%%%%%%%%%%
\subsection{Birman-Schwinger Analysis}
%%%%%%%%%%%%%%%%%%%%%%%%%%%%%%%%%%%%%%
Using the orthonormal basis of~$\sii(M)$ given by~$\{\chi_j\}$,
the free resolvent operator~$R_0(\alpha):=(H_0-\alpha^2)^{-1}$ is
decoupled in the following way
$$
  R_0(\alpha)=
  \sum_{j=1}^\infty
  \chi_{j} \, \left(-\Delta+k_j(\alpha)^2\right)^{-1}
  \, \bar{\chi}_{j},
  \qquad
  k_j(\alpha):=\sqrt{\kappa_j^2-\alpha^2}.
$$
If we are interested in eigenvalues of~$H_\lambda$ below the
lowest transverse mode, we have to consider $\alpha\in[0,\kappa_1)$
and the two-dimensional resolvent in the middle of the expansion
is well defined for any~$j=1,2,\dots\:$. It can be expressed in
terms of Hankel's function -- \cf~\cite[Chap.~I.5]{AGH}. Passing
to Macdonald's functions~$K_0$ by~\cite[9.6.4]{AS}, we arrive at
the following integral kernel formula
\begin{equation}\label{resolvent}
  R_0(x,y,x',y';\alpha)=\frac{1}{2\pi}
  \sum_{j=1}^\infty
  \chi_{j}(y) \,
  K_0\left(k_j(\alpha)|x-x'|\right) \,
  \bar{\chi}_{j}(y').
\end{equation}
Define~$K(\alpha):=|V|^\frac{1}{2}R_0(\alpha)V^\frac{1}{2}$, where
we have employed the usual sign convention,
$V^\frac{1}{2}:=|V|^\frac{1}{2}\sgn V$. According to
Birman-Schwinger principle
-- \cf~\cite{SiQF} --
the function $\alpha(\lambda)^2\equiv E(\lambda)$ is an eigenvalue
of~$H_\lambda$ if and only if the operator~$\lambda K(\alpha)$ has
the eigenvalue~$-1$, \ie,
\begin{equation}\label{BS}
  \alpha^2\in\sigma_\mathrm{disc}(H_\lambda)
  \Longleftrightarrow
  -1\in\sigma_\mathrm{disc}(\lambda K(\alpha)).
\end{equation}
The first term in the expansion~(\ref{resolvent}) referring to
$j=1$ has a singularity at~$\alpha=\kappa_1$, and as usual in such
problems we have to single it out. For this purpose we first
introduce the following decomposition.
\begin{lemma}\label{Interpolation}
There are real-analytic functions~$f$ and~$g$ such that
$$
  \forall u\in(0,\infty):
  \quad
  K_0(u)=f(u) \ln u +g(u)
$$
and
\begin{enumerate}
\item[\emph{(i)}]\
    $f(u)=-1+\mathcal{O}(u^2)$, \
        $g(u)=(\ln 2-\Euler)+\mathcal{O}(u^2)$
    \quad as \ $u\to 0+$,
\item[\emph{(ii)}]\
    $\exists\, C_1>0\; \forall u\in (0,\infty):\
    \max\{f(u),g(u)\}\leq C_1 e^{-u}$,
\end{enumerate}
where~$\Euler$ denotes Euler's constant.
\end{lemma}
\begin{PROOF}
Around the origin a good choice to approximate the Macdonald
function~$K_0$ would be $f:=-I_0$, where~$I_0$ is the other
modified Bessel function~\cite[9.6]{AS} but it has a bad behaviour
at large distances. Hence we use an interpolation, for instance,
\begin{eqnarray*}
  f(u) & := & -e^{-u^2} I_0(u)
              -(1-e^{-u^2}) \, K_0(u) \\
  g(u) & := & e^{-u^2} I_0(u) \, \ln u
              +\big[1+(1-e^{-u^2}) \ln u\big]\,K_0(u).
\end{eqnarray*}
Using the relation~\cite[9.6.13]{AS}, we obtain the behaviour at
the origin~(i). On the other hand, it follows by~\cite[9.7.2]{AS}
that~$f$ and~$g$ have a faster-than-exponential decay at
infinity, which gives together with~(i) that there is a
positive~$C_1$ such that~(ii) holds.
\end{PROOF}

We will thus use the decomposition $K(\alpha)=L_\alpha+M_\alpha$,
where
$$
  L_\alpha(x,y,x',y'):=-\frac{1}{2\pi} \, |V(x,y)|^\frac{1}{2} \,
  \chi_1(y) \, \ln k_1(\alpha) \, \chi_1(y') \, V(x',y')^\frac{1}{2}
$$
contains the singularity and the regular~$M_\alpha$ splits
into two parts again, $M_\alpha=A_\alpha+B_\alpha$.
The operator $B_\alpha$ is given by the projection of the resolvent
on higher transverse modes, \ie,
$B_\alpha:=|V|^\frac{1}{2}R_0^\bot(\alpha)V^\frac{1}{2}$ with
\begin{equation}\label{higherResolvent}
  R_0^\bot(\alpha):=
  \sum_{j=2}^\infty
  \chi_{j} \, \left(-\Delta+k_j(\alpha)^2\right)^{-1}
  \, \bar{\chi}_{j},
\end{equation}
and the kernel of the remaining term is therefore
\begin{eqnarray*}
\lefteqn{A_\alpha(x,y,x',y')} \\
&&  :=\frac{1}{2\pi} \, |V(x,y)|^\frac{1}{2} \,
  \chi_1(y) \,
  \Big(K_0(k_1(\alpha)|x-x'|)+\ln k_1(\alpha)\Big)
  \, \chi_1(y') \, V(x',y')^\frac{1}{2}.
\end{eqnarray*}
We note that~$M_\alpha$ is by definition well defined for~$\alpha=\kappa_1$.
In particular,
$$
  A_{\kappa_1}(x,y,x',y')=-\frac{1}{2\pi} \,
  |V(x,y)|^\frac{1}{2} \, \chi_1(y)
  \left(\Euler+\ln\frac{|x-x'|}{2}\right) \,
  \chi_1(y') V(x',y')^\frac{1}{2}.
$$
Furthermore, we have the following lemma. Since its proof is
purely technical, we postpone it to Section~\ref{Sec.Lemma} below.
\begin{lemma}\label{Lemma}
Assume~\emph{\AssA}. Then there are positive~$C_2,C_3$ and~$C_4$
such that
\begin{enumerate}
\item[\emph{(i)}]
    $\forall\alpha\in[0,\kappa_1]:\quad \|M_\alpha\|<C_2$,
\item[\emph{(ii)}]
    $\|M_{\alpha}-M_{\kappa_1}\|\leq C_3 \lambda^\gamma$,
    with~$\gamma:=\min\{1,\delta/2\}$,
\item[\emph{(iii)}]
    $\left\|\frac{dM_{\alpha(w)}}{dw}\right\|<C_4 |w|^{-1}$
    for~$\lambda$ sufficiently small,
    where~$w:=(\ln k_1(\alpha))^{-1}$.
\end{enumerate}
\end{lemma}\noindent
We recall that Proposition~\ref{Proposition} yields
$\alpha^2\to\kappa_1^2-$ as~$\lambda\to 0+$, and consequently,
$k_1(\alpha)\to 0+$. Hence the auxiliary variable~$w$ is well
defined and negative for~$\lambda$ small enough, and $w\to 0-$
as~$\lambda\to 0+$.

By the Birman-Schwinger principle~(\ref{BS}) eigenvalues
of~$H_\lambda$ correspond to singularities of the operator
$(I+\lambda K(\alpha))^{-1}$ which we can equivalently express as
$$
  \left(I+\lambda K(\alpha)\right)^{-1}
  =\left[I+\lambda(I+\lambda M_\alpha)^{-1}L_\alpha\right]^{-1}
  (I+\lambda M_\alpha)^{-1}.
$$
Since~$M_\alpha$ is bounded independently of~$\alpha$ due
to~Lemma~\ref{Lemma}(i), we have $\|\lambda M_\alpha\|<1$ for all
sufficiently small~$\lambda$, and therefore the second term on the
right hand side of the last relation is a bounded operator. On the
other hand, $\lambda(I+\lambda M_\alpha)^{-1}L_\alpha$ is a
rank-one operator of the form~$(\psi,\cdot)\varphi$, where
\begin{eqnarray*}
  \psi(x,y) & := & -\frac{\lambda}{2\pi} \, \ln k_1(\alpha) \,
  V(x,y)^\frac{1}{2} \, \chi_1(y), \\
  \varphi(x,y) & := & [(I+\lambda M_\alpha)^{-1}|V|^\frac{1}{2}
  \chi_1](x,y),
\end{eqnarray*}
so it has just one eigenvalue which is~$(\psi,\varphi)$.
The requirement that the latter equals~$-1$ yields the implicit equation
\begin{equation}\label{implicit}
  w=F(\lambda,w), \qquad
  F(\lambda,w):=\frac{\lambda}{2\pi}\left(V^\frac{1}{2} \chi_1,
  \left(I+\lambda M_{\alpha(w)}\right)^{-1}|V|^\frac{1}{2}\chi_1\right),
\end{equation}
where we use the auxiliary variable~$w$ defined in Lemma~\ref{Lemma}
which determines the energy via $\alpha^2=\kappa_1^2-e^{2w^{-1}}$.
Solving~(\ref{implicit}), we arrive at the main result of this section.
\begin{thm}\label{Thm.Expansion}
Assume~\emph{$\langle\textsf{a0--3}\rangle$} and exclude the
trivial case, $V\equiv 0$. Then $H_\lambda$ has for sufficiently
small~$\lambda>0$ exactly one eigenvalue $E(\lambda)$ if and only
if
\begin{equation}\label{Condition.V<0}
  \int_{\Real^2} V_{11}(x) \, dx \leq 0,
\end{equation}
and in this case we can express it as
$
  E(\lambda)=\kappa_1^2-e^{2 w(\lambda)^{-1}},
$
where~$w(\lambda)$ has the following asymptotic expansion:
\begin{eqnarray}\label{expansion}
  w(\lambda)
&=&\frac{\lambda}{2\pi} \int_{\Real^2} V_{11}(x) \, dx \nonumber \\
&+&\left(\frac{\lambda}{2\pi}\right)^2 \Bigg\{
  \int_{\Real^2\times\Real^2} V_{11}(x) \left(\Euler+\ln\frac{|x-x'|}{2}\right)
  V_{11}(x') \, dx\,dx' \nonumber \\
&&-\sum_{j=2}^\infty \int_{\Real^2\times\Real^2}
  V_{1j}(x) \, K_0(k_j(\kappa_1)|x-x'|) \, V_{j1}(x') \, dx\,dx'
  \Bigg\} \nonumber \\
&+&\mathcal{O}(\lambda^{2+\gamma})
\end{eqnarray}
with~$\gamma:=\min\{1,\delta/2\}$.
\end{thm}
\begin{PROOF}
Inserting the identity
$$
  (I+\lambda M_\alpha)^{-1}=I-\lambda M_{\kappa_1}
  -\lambda (M_\alpha-M_{\kappa_1})
  +\lambda^2 M_\alpha^2 (I+\lambda M_\alpha)^{-1},
$$
into~(\ref{implicit}) and employing Lemma~\ref{Lemma}(i-ii), we
get the asymptotic expansion~(\ref{expansion}).
Note that its coefficients are well defined owing to~\Assiii\/;
in particular, for the second-order terms it is true since
$(V^\frac{1}{2}\chi_1,M_{\kappa_1}|V|^\frac{1}{2}\chi_1)$
estimated by the Schwarz inequality
is finite because of~\Assiii\/ and Lemma~\ref{Lemma}(i). The sufficient and
necessary condition~(\ref{Condition.V<0}) follows from the fact
that~$E(\lambda)$ converges to~$\kappa_1^2$ as~$\lambda\to 0+$
because of Proposition~\ref{Proposition}, which corresponds to the
situation that~$w$ goes to zero assuming \emph{negative} values.
In view of~(\ref{expansion}) it is evident that this is the case
if $\int V_{11}$ is strictly negative. We want to show that~$w$ is
negative for small~$\lambda$ also if the first term in the
expansion vanishes. Suppose first that the potential
projections~$V_{jj'}$ belong to~$\sii(\Real^2)$. Then we have
\begin{eqnarray}\label{FourierTrick}
\lefteqn{\int_{\Real^2\times\Real^2} V_{1j}(x) \,
  K_0(k_j(\kappa_1)|x-x'|) \, V_{j1}(x') \, dx\,dx'} \nonumber \\
&& =\left(V_{1j},K_0*V_{j1}\right)
   = \left(\widehat{V_{1j}},\widehat{K_0*V_{j1}}\right) \nonumber \\
&& =2\pi \left(\widehat{V_{1j}},\widehat{K_0}
\widehat{V_{j1}}\right)
  =(2\pi)^2 \int_{\Real^2} \frac{|\widehat{V_{1j}}(\omega)|^2}
  {|\omega|^2+\kappa_j^2-\kappa_1^2} \, d\omega>0,
\end{eqnarray}
because~$(2\pi)^{-1}K_0(k_j(\kappa_1)|\cdot|)$ is the Green
function of~$-\Delta+k_j(\kappa_1)^2$. At the same time, by
Lemma~\ref{Interpolation}(i) and the fact that we
deal with the case~\mbox{$\int V_{00}=0$} now,
\begin{eqnarray}\label{EpsilonTrick}
\lefteqn{-\int_{\Real^2\times\Real^2}
  V_{11}(x) \, \left(\Euler+\ln\frac{|x-x'|}{2}\right) \,
  V_{11}(x') \, dx\,dx'} \nonumber \\
&& = \lim_{\varepsilon\to 0+}\int_{\Real^2\times\Real^2}
  V_{11}(x) \, \left(K_0(\varepsilon|x-x'|)+\ln\varepsilon\right) \,
  V_{11}(x') \, dx\,dx' \nonumber \\
&& =\lim_{\varepsilon\to 0+}\int_{\Real^2\times\Real^2}
  V_{11}(x) \, K_0(\varepsilon|x-x'|) \,
  V_{11}(x') \, dx\,dx'
\end{eqnarray}
is also positive, which follows by the consecutive use
of the Fourier transform trick~(\ref{FourierTrick}).
For a general~$V_{jj'}\not\in\sii(\Real^2)$
we can approximate the potential by the cut-off functions
$
  V^N:=\chi_{[-N,N]^2} \sgn V \min\{|V|,N\},
$
where~$\chi_\mathcal{A}$ denotes the characteristic function of a
set~$\mathcal{A}$. Then the expressions in the first line
of~(\ref{FourierTrick}) and the last line of~(\ref{EpsilonTrick})
are approximated by sequences whose elements are positive by the
above argument. Using the dominated convergence and the absolute
continuity of the Lebesgue integral we find that the limits exist,
and of course they are positive again. Together we get that the
second-order term in the expansion~(\ref{expansion}) is negative.

It remains to check that~(\ref{expansion}) is the only solution
of~(\ref{implicit}) for~$\lambda$ small. This will be true if we
prove it for a non-positive~$V$ since it represents a more
attractive interaction; we thus replace~$V$ by~$-|V|$ in the
expression for~$F$. Using the Schwarz and triangle inequalities
together with Lemma~\ref{Lemma}(i),~(iii), we arrive at the
estimate
\begin{eqnarray*}
\lefteqn{\left|\frac{\partial F}{\partial w}(\lambda,w)\right|} \\
&& =\frac{\lambda}{2\pi} \left|
  \left( |V|^\frac{1}{2}\chi_1,
  (I+\lambda M_{\alpha(w)})^{-1}
  \lambda\frac{dM_{\alpha(w)}}{dw}
  (I+\lambda M_{\alpha(w)})^{-1}
  |V|^\frac{1}{2}\chi_1 \right) \right| \\
&&\leq \frac{\lambda^2}{2\pi}
  \left\||V|^\frac{1}{2}\chi_1\right\|_2^2
  \left\|\left(I+\lambda M_{\alpha(w)}\right)^{-1}\right\|^2
  \left\|\frac{dM_{\alpha(w)}}{dw}\right\| \\
&&\leq \frac{\lambda^2}{2\pi}\,(1-\lambda C_2)^{-2} \frac{C_4}{|w|}
  \left\||V|_{11}\right\|_1^2,
\end{eqnarray*}
which holds for~$\lambda$ sufficiently small and strictly less
than~$C_2^{-1}$. The norm of the potential is finite by
assumption~\Assiii\/. Excluding the trivial case \mbox{$V\equiv
0$} we note that there exists a~$c'>0$ such that the inequality
\mbox{$|w|^{-1}\leq c' \lambda^{-1}$} is valid for any
solution~$w$ of the implicit equation~(\ref{implicit})
and~$\lambda$ small enough. So there is a~$C_5>0$ such that the
partial derivative of~$F$ w.r.t.~$w$ is bounded by~$C_5\lambda$
for all sufficiently small~$\lambda$. Since any two
solutions~$w_1,w_2$ of the equation~$w=F(\lambda,w)$ have to
fulfill
$$
  |w_2-w_1|=\left|\int_{w_1}^{w_2}\frac{\partial F}{\partial w}\,dw \right|
  \leq \left|\int_{w_1}^{w_2}
  \left|\frac{\partial F}{\partial w}\right|\,dw \right|
  \leq C_5 \lambda\,|w_2-w_1|,
$$
the uniqueness is ensured for~$\lambda<C_5^{-1}$.
\end{PROOF}
%

%%%%%%%%%%%%%%%%%%%%%%%%%%%%%%%%%%%%%%%
\subsection{Proof of Lemma~\ref{Lemma}}\label{Sec.Lemma}
%%%%%%%%%%%%%%%%%%%%%%%%%%%%%%%%%%%%%%%
A reader not interested in the following technical analysis of the
operator~$M_\alpha$ may skip this subsection. We recall
that~$M_\alpha$ is given by the sum of~$A_\alpha$ and~$B_\alpha$,
which are of a different nature. To prove the assertions
of~Lemma~\ref{Lemma} we consider each of the operators separately.

%%%%%%%%%%%%%%%%%%%%%%%%%%%%%%%%%%%%%%
\subsubsection{Analysis of $A_\alpha$}
%%%%%%%%%%%%%%%%%%%%%%%%%%%%%%%%%%%%%%
We show first that this operator is of the Hilbert-Schmidt class
for~$\alpha=\kappa_1$.
\begin{lemma}\label{A-Bound}
Asuume \emph{\Assii}, \emph{\Assiii}, then $A_{\kappa_1}$ is a
Hilbert-Schmidt operator.
\end{lemma}
\begin{PROOF}
Let us compute the HS-norm
$$
  \|A_{\kappa_1}\|_{HS}^2=\frac{1}{4\pi^2}
  \int_{\Real^2} \int_{\Real^2}|V|_{11}(x)
  \left|\Euler+\ln\frac{|x-x'|}{2}\right|^2 |V|_{11}(x')\,dx\,dx'.
$$
It can be estimated by a sum of two integrals. The first one will
be finite if $|V|_{11}\in\si(\Real^2)$ which is true by~\Assiii\/,
so it remains to check that the integral
$$
  J:=\int_{\Real^2\times\Real^2}
  |V|_{11}(x) \, \ln^2|x-x'| \, |V|_{11}(x')\,dx\,dx'
$$
is finite. To estimate it, we divide the region of integration
into two parts: if $|x-x'|\geq 1$, then for any~$\delta>0$ there
is a $C_\delta>0$ such that
$$
  \ln^2|x-x'| \leq \ln^2(|x|+|x'|) \leq \ln^2(1+|x|)(1+|x'|)
  \leq C_\delta (1+|x|^\delta) (1+|x'|^\delta)
$$
and the contribution to~$J$ is thus finite because of~\Assiii. On
the other hand, for $|x-x'|<1$ we use the H\"older and Young
inequalities
\begin{eqnarray*}
  J
&=&\int_{\Real^2} dx \, |V|_{11}(x) \int_{\Real^2} dx'
  \chi_{[0,1]}(|x-x'|) \, \ln^2|x-x'| \, |V|_{11}(x') \\
&\leq& \||V|_{11}\|_{1+\delta} \,
  \|\chi_{[0,1]}\ln^2*\,|V|_{11}\|_{1+\delta^{-1}} \\
&\leq& \||V|_{11}\|_{1+\delta}^2 \,
  \|\chi_{[0,1]}\ln\|_r
  =\Gamma(r+1)^{\frac{1}{r}} \, \||V|_{11}\|_{1+\delta}^2\,,
\end{eqnarray*}
where~$r:=(1+\delta^{-1})/2$, which yields a finite value owing
to~\Assii.
\end{PROOF}

In a similar way we can estimate the HS-norm
of~$A_\alpha-A_{\kappa_1}$.
\begin{lemma}\label{A-Continuity}
Assume \emph{\Assiii}, then $\|A_\alpha-A_{\kappa_1}\|_{HS} \leq
C_6\,k_1(\alpha)^\delta$ holds with a positive constant~$C_6$
independent of~$\alpha$.
\end{lemma}
\begin{PROOF}
Using the decomposition of Lemma~\ref{Interpolation},
\begin{eqnarray*}
\lefteqn{(A_\alpha-A_{\kappa_1})(x,y,x',y')} \\
&& \qquad\qquad
   =|V(x,y)|^\frac{1}{2} \, \chi_1(y)\,
   \Big\{[f(k_1(\alpha)|x-x'|)-f(0)] \ln k_1(\alpha)|x-x'|\\
&& \qquad\qquad
   \phantom{=} +g(k_1(\alpha)|x-x'|)-g(0)\Big\}
   \,\chi_1(y') \, V(x',y')^\frac{1}{2},
\end{eqnarray*}
which yields the estimate $\|A_\alpha-A_{\kappa_1}\|_{HS}^2 \leq
2\,(J_1^2+J_2^2)$, where
$$
  J_\ell:=\int_{\Real^2\times\Real^2}
  |V|_{11}(x) \, h_\ell(k_1(\alpha)|x-x'|) \, |V|_{11}(x') \, dx\,dx',
  \quad\ell=1,2\,,
$$
with~$h_1(u):=[(f(u)-f(0))\ln u]^2$ and~$h_2(u):=[g(u)-g(0)]^2$.
Since these functions are bounded, both integrals are finite under
the assumption $|V|_{11}\in\si(\Real^2)$. Consequently,
$\|A_\alpha-A_{\kappa_1}\|_{HS}$ has a bound independent
of~$\alpha$. However, we want to prove in addition that~$A_\alpha$
converges in HS-norm to~$A_{\kappa_1}$, \ie, the stated assertion.
By virtue of~Lemma~\ref{Interpolation}, we have for any~$\delta\in(0,2)$
the rough bounds $h_\ell(u)\leq c_\ell u^\delta$ for some
constants~$c_\ell>0$. This bound together with $|x-x'|^\delta\leq
\max\{1,2^{\delta-1}\}\,(|x|^\delta+|x'|^\delta)$
and~$|x|^\delta+|x'|^\delta\leq (1+|x|^\delta)(1+|x'|^\delta)$,
yields the estimate
$$
  J_\ell \leq \tilde{C}_6 \,
  k_1(\alpha)^\delta
  \left(\int_{\Real^2}|V|_{11}(x) \, (1+|x|^\delta) \, dx\right)^2,
$$
where $\tilde{C}_6:=\max\{1,2^{\delta-1}\} \max\{c_1,c_2\}$. Since
the integral is finite by~\Assiii\/, we arrive at the sought
result.
\end{PROOF}
It is now easy to check that the two preceding lemmas imply the
claims~(i) and~(ii) of Lemma~\ref{Lemma} for the
operator~$A_\alpha$, since $\|\cdot\|\leq\|\cdot\|_{HS}$
and~$k_1(\alpha)=\mathcal{O}(\lambda^\frac{1}{2})$ as it follows
from Proposition~\ref{Proposition}.

It is also easy to see that~$\alpha\mapsto A_\alpha$ considered as
the operator-valued function is real-analytic in~$[0,\kappa_1)$.
It cannot be analytically continued to an open interval containing
~$\kappa_1$, however, if we consider instead the function
$w\mapsto A_{\alpha(w)}$ with the auxiliary variable
$w=(\ln k_1(\alpha))^{-1}$, this one can be continued to a complex region
that includes~$w=0$, which is the point of our interest because it is
obtained in the limit~$\lambda\to 0$. We use this fact to estimate
the norm of the derivative w.r.t.~$w$, which establishes the
claim~(iii) of Lemma~\ref{Lemma} for the operator~$A_{\alpha(w)}$.
\begin{lemma}\label{dA-Bound}
Assume~\emph{\Assii}\/ and~\emph{\Assiii}, then for sufficiently
small~$w$,
$$
  \left\|\frac{dA_{\alpha(w)}}{dw}\right\| < \frac{C_7}{|w|}
  \quad\textrm{with some}\quad C_7>0.
$$
\end{lemma}
\begin{PROOF}
Let us choose the contour $\Gamma:= \{z=w+w e^{it}\,:\;
t\in[0,2\pi)\,\}$ in the complex plane, where $w$ is supposed to
be negative and small enough so that~$A_{\alpha(w)}$ is analytic
in the interior of the curve. We can use the Cauchy integral
formula to estimate the complex derivative
$$
  \left\|\frac{dA_{\alpha(w)}}{dw}\right\|
  =\left\|\frac{1}{2\pi i}
  \oint_\Gamma \frac{A_{\alpha(z)}}{(z-w)^2}\,dz \right\|
  \leq \frac{\sup_{z\in\Gamma}\|A_{\alpha(z)}\|}{|w|}.
$$
It is straightforward to modify the proof of
Lemma~\ref{A-Continuity} (including the technical
Lemma~\ref{Interpolation}) in order to check the continuity
of~$A_\alpha$ in~$\kappa_1$ w.r.t. to the HS-norm also for complex
values of~$\alpha$. This yields the desired result.
\end{PROOF}
%

%%%%%%%%%%%%%%%%%%%%%%%%%%%%%%%%%%%%%%
\subsubsection{Analysis of $B_\alpha$}
%%%%%%%%%%%%%%%%%%%%%%%%%%%%%%%%%%%%%%
Mimicking~\cite[Lemma~2.2]{BGRS} and~\cite[Lemma~2.3]{BEGK} we
denote as $\mathcal{H}_1\subset\mathcal{H}$ the subspace
$\sii(\Real^2)\otimes \{\chi_1\}$.
Let~$\mathcal{P}_1$ be the corresponding projection
and \mbox{$\proj:=I-\mathcal{P}_1$}, then we can write
$R_0^\bot(\alpha)$ defined in~(\ref{higherResolvent})
as $\proj R_0(\alpha) \proj$.
\begin{lemma}\label{B-Bound}
Assume~\emph{\Assi}, then $\alpha\mapsto B_\alpha$ is uniformly
bounded in the operator-norm topology on the
interval~$[0,\kappa_2-\varepsilon]$ for
any~$\varepsilon\in(0,\kappa_2]$.
\end{lemma}
\begin{PROOF}
Since the lowest point in the spectrum of
$H_0\proj\upharpoonright\proj\mathcal{H}$ is~$\kappa_2^2$, the
operator-valued function~$R_0^\bot(\cdot)$ has an analytic
continuation into the region $\{\alpha\in\Com\, |\,
\alpha^2\in\Com\setminus\,[\kappa_2^2,\infty)\,\}$. In particular,
this region includes the interval~$[0,\kappa_1]$ actually
considered, where one has the following estimate on the norm
$$
  \left\|R_0^\bot(\alpha)\right\|
  =\sup_{j=2,3,\dots} (\kappa_j^2-\alpha^2)^{-1}
  = (\kappa_2^2-\alpha^2)^{-1} < (\kappa_2^2-\kappa_1^2)^{-1}.
$$
If~$V$ was essentially bounded then this result would be
sufficient for the boundedness of~$B_\alpha$ because
$\|B_\alpha\|\leq\|V^\frac{1}{2}\|_\infty
\|R_0^\bot(\alpha)\|\,\|V^\frac{1}{2}\|_\infty$. In order to
accommodate the extra factors $|V|^\frac{1}{2},V^\frac{1}{2}$
under our weaker assumption~\Assi\/, we introduce the quadratic
form
$$
  b_\alpha(\phi,\psi):=(\phi,B_\alpha\psi)
  = \left(R_0^\bot(\alpha)^\frac{1}{2} \proj |V|^\frac{1}{2} \phi,
  R_0^\bot(\alpha)^\frac{1}{2} \proj V^\frac{1}{2} \psi\right)
$$
with $\alpha$ supposed to be a real number from an
interval~$[0,\kappa_2-\varepsilon]$. To check boundedness of this
form, it is sufficient to verify that
$R_0^\bot(\alpha)^\frac{1}{2} \proj |V|^\frac{1}{2}$ is a bounded
operator, \ie, that $|V|^\frac{1}{2}$ is
$(R_0^\bot(\alpha)^{-\frac{1}{2}}\proj)$-bounded, which is
equivalent to the statement that there exist~$c_1,c_2 \geq 0$ such
that
$$
  \forall\psi\in\Dom Q_0: \quad
  \left\|V^\frac{1}{2}\proj\psi\right\|^2 \leq
  c_1 \left\|\psi\right\|^2
  +c_2 \left\|R_0^\bot(\alpha)^{-\frac{1}{2}}\proj\psi\right\|^2.
$$
However,
{\setlength\arraycolsep{2pt}
\begin{eqnarray*}
  \left\|V^\frac{1}{2}\proj\psi\right\|^2
&=& \left(\proj\psi,|V|\proj\psi\right)
  \leq \left\|\proj\psi\right\| \, \left\|V\proj\psi\right\| \\
&\leq& \left\|\proj\psi\right\| \left(a \left\|\proj\psi\right\|
  + b \left\|H_0^\frac{1}{2}\proj\psi\right\|\right) \\
&\leq& \left(a+\frac{b}{2}\right) \left\|\psi\right\|^2
  +\frac{b}{2} \left\|H_0^\frac{1}{2}\proj\psi\right\|^2.
\end{eqnarray*}
}\noindent
In the first step we have used the Schwarz inequality, in the
second our hypothesis~\Assi, and finally the inequality between
the geometric and arithmetic means together with
$\|\proj\psi\|\leq\|\psi\|$. At the same time, $$
  \left\|R_0^\bot(\alpha)^{-\frac{1}{2}}\proj\psi\right\|^2
  =\left(\proj\psi,(H_0-\alpha^2)\proj\psi\right)
  \geq \left\|H_0^\frac{1}{2}\proj\psi\right\|^2,
$$
so we can identify $c_1:=a+b/2$ and $c_2:=b/2$ which completes
our proof.
\end{PROOF}

This establishes Lemma~\ref{Lemma}(i) for the operator~$B_\alpha$.
Since~$k_1(\alpha)^2\leq c\lambda$ owing to
Proposition~\ref{Proposition}, the property~(ii) is included in
the following result.
\begin{lemma}\label{B-Continuity}
Assume \emph{\Assi}, then there exists~$C_7>0$ such that
$$
  \left\|B_\alpha-B_{\kappa_1}\right\| \leq C_7\,k_1(\alpha)^2.
$$
\end{lemma}
\begin{PROOF}
Using the first resolvent identity~\cite[Thm.~5.13]{Weidmann}, we infer
\begin{multline*}
\|B_\alpha-B_{\kappa_1}\|
  = |\alpha-\kappa_1| \left\| |V|^\frac{1}{2} R_0^\bot(\alpha)
  R_0^\bot(\kappa_1) V^\frac{1}{2} \right\| \\
\leq |\alpha-\kappa_1|
  \left\| |V|^\frac{1}{2} R_0^\bot(\alpha)^\frac{1}{2} \right\|
  \left\|R_0^\bot(\alpha)^\frac{1}{2}\right\|
  \left\|R_0^\bot(\kappa_1)^\frac{1}{2}\right\|
  \left\| R_0^\bot(\kappa_1)^\frac{1}{2} V^\frac{1}{2} \right\|.
\end{multline*}
However, $|\kappa_1-\alpha|\leq\kappa_1^{-1}k_1(\alpha)^2$ and it
is clear from the proof of Lemma~\ref{B-Bound} that the remaining
factors at the r.h.s. of the above estimate are finite.
\end{PROOF}

Similarly to the operator~$A_\alpha$, we need also a bound on the
derivative of~$B_{\alpha(w)}$ w.r.t.~$w$. However, the situation
in the present case is simpler because~$R_0^\bot(\alpha)$ itself
is analytic in an open complex set containing~$\alpha=\kappa_1$.
\begin{lemma}\label{dB-Bound}
Assume \emph{\Assi}, then for sufficiently small~$w$,
$$
  \left\|\frac{dB_{\alpha(w)}}{dw}\right\| \leq C_8
  \quad\textrm{with some}\quad C_8>0.
$$
\end{lemma}
\begin{PROOF}
Since
$$
  \frac{dB_{\alpha(w)}}{dw} = \frac{dB_\alpha}{d\alpha} \frac{d\alpha}{dw}
  = 2 \frac{e^{2w^{-1}}}{w^2} \,
  |V|^\frac{1}{2} R_0^\bot(\alpha(w))^2 V^\frac{1}{2}
$$
and the prefactor function~$w^{-2}e^{2w^{-1}}$
is uniformly bounded in~$(-\infty,0)$,
we can employ the Schwarz inequality to get the estimate
$$
  \left\| |V|^\frac{1}{2} R_0^\bot(\alpha)^2 V^\frac{1}{2} \right\|
  \leq \left\||V|^\frac{1}{2} R_0^\bot(\alpha)^\frac{1}{2}\right\|
  \left\|R_0^\bot(\alpha)\right\|
  \left\|R_0^\bot(\alpha) V^\frac{1}{2}\right\|.
$$
It is clear from the proof of Lemma~\ref{B-Bound} that the right
hand side of this inequality is uniformly bounded
in~$[0,\kappa_2-\varepsilon]$ for any~$\varepsilon\in(0,\kappa_2]$.
The former interval contains a neighbourhood of~$\kappa_1$
in which~$w$ is well defined.
\end{PROOF}
\noindent
This yields the remaining assertion~(iii) of
Lemma~\ref{Lemma} for~$|w|<1$.

\setcounter{equation}{0}
%%%%%%%%%%%%%%%%%%%%%%%%%%%%%%
\section{Mildly Curved Layers}\label{Sec.BS}
%%%%%%%%%%%%%%%%%%%%%%%%%%%%%%
Our strategy in proving the ground-state asymptotic expansion in
mildly curved layers~$\Omega_\varepsilon$ will be to estimate the
corresponding Hamiltonian~$-\Delta_D^{\Omega_\varepsilon}$ by an
operator of the form~$-\Delta-\Delta_D^I+\varepsilon V$ and to
apply Theorem~\ref{Thm.Expansion} to the latter. Here~$-\Delta$ is
the Laplacian in the plane, $-\Delta_D^I$ is the transverse
operator which is the particular case of~$-\Delta_D^M$ discussed
in the preceding section, and~$V$ is an effective potential given
by curvatures of~$\Sigma_\varepsilon$.

%%%%%%%%%%%%%%%%%%%%%%%%%
\subsection{The Geometry}
%%%%%%%%%%%%%%%%%%%%%%%%%
The family of metric tensors for the surfaces given
by~(\ref{surface}) has the form
$$
  g_{\mu\nu}(\varepsilon)=\delta_{\mu\nu}+\varepsilon^2 \eta_{\mu\nu}
  \qquad\textrm{with}\qquad
  (\eta_{\mu\nu}):=
  \begin{pmatrix}
    {f_{,1}}^2    & f_{,1} f_{,2} \\
    f_{,1} f_{,2} & {f_{,2}}^2
  \end{pmatrix},
$$
where the symbol $\delta_{\mu\nu}$ (as well as $\delta^{\mu\nu}$)
has to be understood as the identity matrix. Since
$\det(\eta_{\mu\nu})=0$ we get immediately
\begin{equation}\label{WCLg}
  g(\varepsilon):=\det(g_{\mu\nu})=1+\varepsilon^2\tr(\eta_{\mu\nu})
  =1+\varepsilon^2({f_{,1}}^2+{f_{,2}}^2)
\end{equation}
together with the expression for the inverse matrix
$$
  g^{\mu\nu}(\varepsilon)
  =g(\varepsilon)^{-1}
  (\delta^{\mu\nu}+\varepsilon^2 \tilde{\eta}^{\mu\nu})
  \qquad\textrm{with}\qquad
  (\tilde{\eta}^{\mu\nu}):=
  \begin{pmatrix}
    {f_{,2}}^2     & -f_{,1} f_{,2} \\
    -f_{,1} f_{,2} & {f_{,1}}^2
  \end{pmatrix}.
$$
In particular, the Jacobian~$g^\frac{1}{2}$ defines through
$d\Sigma_\varepsilon:=g^\frac{1}{2} dx$ the invariant surface
element. We will suppose that the matrix function~$\eta_{\mu\nu}$
is bounded. Since its eigenvalues are~$0$ and
\mbox{${f_{,1}}^2+{f_{,2}}^2$}, this will be true provided the
second eigenvalue is a bounded function in~$\Real^2$, \ie, if we
adopt the assumption~\AssDi. Denote the bound of~$\eta_{\mu\nu}$
by~$\eta_\infty$. Then it follows that $g_{\mu\nu}(\varepsilon)$
is uniformly elliptic for sufficiently small~$\varepsilon$ because
\begin{equation}\label{c+-}
  c_-\delta_{\mu\nu}
  \leq g_{\mu\nu}(\varepsilon) \leq
  c_+\delta_{\mu\nu}
  \qquad\textrm{with}\qquad
  c_\pm:=1\pm\varepsilon^2\eta_\infty.
\end{equation}
At the same time, the vector
$n(\varepsilon)=g(\varepsilon)^{-\frac{1}{2}}
(-\varepsilon f_{,1},-\varepsilon f_{,2},1)$
represents the surface normal and therefore the second fundamental
form is given by
$$
  h_{\mu\nu}(\varepsilon)
  =\varepsilon g(\varepsilon)^{-\frac{1}{2}}\theta_{\mu\nu}
  \qquad\textrm{with}\quad
  (\theta_{\mu\nu}):=
  \begin{pmatrix}
    f_{,11} & f_{,12} \\
    f_{,21} & f_{,22}
  \end{pmatrix}.
$$
We can construct now
the Weingarten tensor \cite[Def.~3.3.4 \&~Prop.~3.5.5]{Kli}
$$
  h_\mu^{\ \nu}(\varepsilon):=h_{\mu\rho}g^{\rho\nu}
  =\varepsilon g(\varepsilon)^{-\frac{3}{2}}
  \left(\theta_{\mu\rho}\delta^{\rho\nu}
  +\varepsilon^2\theta_{\mu\rho}\tilde{\eta}^{\rho\nu}\right),
$$
which determines respectively the Gauss curvature~$K$ and the mean
curvature~$M$ of the surface $\Sigma_\varepsilon$:
\begin{multline}\label{km}
\begin{aligned}
  K(\varepsilon)
&=\varepsilon^2 g(\varepsilon)^{-2} k_0
&\ \textrm{with}\quad &
  k_0:=\det(\theta_{\mu\nu})=f_{,11}f_{,22}-{f_{,12}}^2 , \\
  M(\varepsilon)
&=\varepsilon g(\varepsilon)^{-\frac{3}{2}}
  \left(m_0+\varepsilon^2 m_1\right)
&\ \textrm{with}\quad &
  m_0:=\frac{1}{2}\tr(\theta_{\mu\nu})
  =\frac{1}{2} \left(f_{,11}+f_{,22}\right)
\end{aligned} \\
  \textrm{and}\quad
  m_1:=\frac{1}{2}\tr(\theta_{\mu\rho}\tilde{\eta}^{\rho\nu})
  =\frac{1}{2}
  \left({f_{,1}}^2 f_{,22}+{f_{,2}}^2 f_{,11}-2f_{,1} f_{,2} f_{,12}\right).
\end{multline}
Since we are interested in the case when~$\Sigma_\varepsilon$ is
asymptotically planar~\cite{DEK1}, \ie, when $K,M\to 0$
as~$|x|\to\infty$, we assume~\AssDii.

It is clear from the definition~(\ref{layer}) that the metric
tensor of the layer~$\Omega_\varepsilon$ (as a manifold with a
boundary in~$\Real^3$) has the block form
\begin{equation}\label{matrixG}
  (G_{ij})=
\begin{pmatrix}
  (G_{\mu\nu}) & 0 \\
  0            & 1 \\
\end{pmatrix}
  \quad\textrm{with}\quad
  G_{\nu\mu}=
  (\delta_\nu^\sigma-u h_\nu^{\ \sigma})
  (\delta_\sigma^\rho-u h_\sigma^{\ \rho})
  g_{\rho\mu},
\end{equation}
and thus~$G:=\det(G_{ij})=g(1-2Mu+Ku^2)^2$. Following~\cite{DEK1},
one should make sure that the layer mapping~$\layer$ is a
diffeomorphism. However, this is automatically fulfilled for an
arbitrary~$a$ provided~$\varepsilon$ is small enough and the layer
is built over surfaces of the special form~(\ref{surface}). At the
same time, $G_{\mu\nu}$ can be estimated by the surface metric,
\begin{equation}\label{C+-}
  C_- g_{\mu\nu}\leq G_{\mu\nu}\leq C_+ g_{\mu\nu}
  \qquad\textrm{with}\qquad
  C_\pm:=\left(1\pm a\rho_m^{-1}\right)^2,
\end{equation}
where
$\rho_m^{-1}:=\max\left\{\|k_1\|_\infty,\|k_2\|_\infty\right\}
=\mathcal{O}(\varepsilon)$
and~$k_1,k_2$ are the principal curvatures of~$\Sigma_\varepsilon$,
\ie, the eigenvalues of~$h_\mu^{\ \nu}$.

%%%%%%%%%%%%%%%%%%%%%%%%%%%%
\subsection{The Hamiltonian}
%%%%%%%%%%%%%%%%%%%%%%%%%%%%
In the beginning we have identified the particle Hamiltonian
with~$-\Delta_D^{\Omega_\varepsilon}$. We want to replace it by a
Schr\"odinger-type operator which would allow us to employ the
result of the previous section. This is achieved by means of the
unitary transformation
$$
  U:\sii(\Omega_\varepsilon)\to\sii(\Omega_0,g^\frac{1}{2}dx\,du):
  \{\psi\mapsto U\psi:=(1-2Mu+Ku^2)^\frac{1}{2}\,\psi\circ\!\layer\,\},
$$
which leads to the unitarily equivalent operator
\begin{align*}
H&:=U (-\Delta_D^{\Omega_\varepsilon}) U^{-1}
  =-g^{-\frac{1}{2}}\partial_i g^{\frac{1}{2}}G^{ij}\partial_j+V \\
V&=g^{-\frac{1}{2}}(g^\frac{1}{2}G^{ij} J_{,j})_{,i}
  +J_{,i}G^{ij}J_{,j}
  \qquad\textrm{with}\quad
  J:=\ln\sqrt{1-2Mu+Ku^2}.
\end{align*}
It makes sense since we suppose that the surface
is~$\Smooth^4$-smooth. Using the block form~(\ref{matrixG})
of~$G_{ij}$, we can split~$H$ into a sum, $H=H_1+H_2$. For the
first operator which is given by the part of~$H$ where one sums
over the Greek indices (referring to the longitudinal coordinates)
we have owing to~(\ref{C+-}) the estimate
\begin{equation}\label{H1-bound}
  (C_-/C_+^2) \left(-\Delta_g+v_1\right)
  \leq H_1 \leq
  (C_+/C_-^2) \left(-\Delta_g+v_1\right)
\end{equation}
which holds in~$\sii(\Real^2\times I,g^\frac{1}{2}dx\,du)$ in the
form sense. Here~$-\Delta_g$ denotes the surface Laplace-Beltrami
operator which can be written in the component form as
$-g^{-\frac{1}{2}}\partial_\mu
g^{\frac{1}{2}}g^{\mu\nu}\partial_\nu$, and
\begin{equation}\label{v1}
  v_1:=-\frac{|u^2\nabla_{\!g}K-2u\nabla_{\!g}M |_g^2}{4(1-2Mu+Ku^2)^2}
  +\frac{u^2\Delta_g K-2u\Delta_g M}{2(1-2Mu+Ku^2)}\,,
\end{equation}
where~$|\cdot|_g$ and~$\nabla_{\!g}$ mean respectively the norm
and the gradient operator induced by the metric~$g_{\mu\nu}$. On
the other hand,
\begin{equation}\label{hatHamiltonian2}
  H_2=-\partial_3^2+V_2\,,
  \qquad
  V_2=\frac{K-M^2}{(1-2Mu+Ku^2)^2}\,,
\end{equation}
where the potential~$V_2$ is attractive because~$K-M^2$ can be
rewritten by means of the principal curvatures
as~$-\frac{1}{4}(k_1\!-\!k_2)^2$.

In the second step one uses the inequalities~(\ref{c+-}), which
are a consequence of~\AssDi\/, in order to get the bounds
$$
  (c_-^2/c_+^3) \left(-\Delta\right)
  \leq -\Delta_g \leq
  (c_+^2/c_-^3) \left(-\Delta\right)
  \qquad\textrm{in}\quad
  \sii(\Real^2,g^\frac{1}{2} dx).
$$
It remains to realize that -- as another consequence of the
uniform ellipticity of the metric -- one can identify
$\sii(\Real^2,g^\frac{1}{2} dx)=\sii(\Real^2)$, and also
$\sii(\Real^2\times I,g^\frac{1}{2}dx\,du)=\sii(\Real^2\times I)$
as sets. If we rescale the longitudinal variable by means of
$x\mapsto\sigma_\pm x$ with~$\sigma_\pm^2:=
c_\mp^3C_\mp^2/(c_\pm^2 C_\pm)$, we obtain finally the above
indicated bounds
\begin{equation}\label{H+-}
  H_- \leq H \leq H_+
  \qquad\textrm{with}\qquad
  H_\pm:=-\Delta-\partial_3^2+\varepsilon V_\pm,
\end{equation}
where
$$
  V_\pm(x,u):=\frac{1}{\varepsilon}
  \left(\frac{C_\pm}{C_\mp^2}\,v_1+V_2\right)
  \left(x/\sigma_\pm,u\right).
$$
We notice that since~$v_1$ and~$V_2$ are~$\varepsilon$-dependent
-- \cf~(\ref{km}) and the explicit formulae for the
potentials~(\ref{v1}) and~(\ref{hatHamiltonian2}) -- the
potentials~$V_\pm$ are well defined for~$\varepsilon=0$.

%%%%%%%%%%%%%%%%%%%%%%%%%%%%
\subsection{The Ground State Asymptotics}\label{Sec.BoundState}
%%%%%%%%%%%%%%%%%%%%%%%%%%%%
Suppose that each of the operators~$H_\pm$ has for all
sufficiently small~$\varepsilon$ just one
eigenvalue~$E_\pm(\varepsilon)$. Since $H_-$ is below bounded, the
minimax principle tells us that the same is true for~$H$ and
\begin{equation}\label{WCL.Squeeze}
  E_-(\varepsilon) \leq E(\varepsilon) \leq E_+(\varepsilon).
\end{equation}
If we identify~$\varepsilon$ with~$\lambda$ of the previous
section, we see that under our assumptions the operators~$H_\pm$
are well suited for application of Theorem~\ref{Thm.Expansion}.

Indeed, the decay requirements~\AssD\/ yield that
$K,M,|\nabla_{\!g}K|_g,|\nabla_{\!g}M|_g$ and~$\Delta_g K$ tend to
zero at infinity, and consequently, $(V_\pm)_{11}\to 0$
as~$|x|\to\infty$. Notice that we do not assume a decay
of~$\Delta_g M$ at infinity which would
require~$f_{,\mu\nu\!\rho\sigma}\to 0$ as~$|x|\to\infty$. The
reason is that the term~$u\Delta_g M$ itself in the
potential~(\ref{v1}) vanishes when projected onto the first
transverse mode~$\chi_1$. One can thus take
$\psi_n(x,u):=\varphi_n(x)\chi_1(u)$, where~$\varphi_n$ with
$\|\varphi_n\|=1$ is a sequence of functions with
increasing supports which move towards infinity as~$n\to\infty$.
If~$\varphi_n$ is properly chosen such~$\psi_n$ form a Weyl sequence
for~$H_\pm$ and any~$\kappa_1^2+\alpha$ with~$\alpha\in\posReal$.
Using then a Neumann bracketing argument, as mentioned in
Section~\ref{res}, in order to show that no~$\kappa_1^2+\alpha$
with~$\alpha<0$ can belong to the essential spectrum, we conclude
that~$\sigma_\mathrm{ess}(H_\pm)=[\kappa_1^2,\infty)$. This
verifies~\AssV\/ of Theorem~\ref{Thm.Expansion}. Furthermore,
using this result together with~(\ref{H+-}) and the minimax
principle, we infer that~$\inf\sigma_\mathrm{ess}(H)=\kappa_1^2$.

The requirement~\Assi\/ holds trivially true because~$V_\pm$ are
essentially bounded due to the surface regularity assumption
and~\AssD. Under \AssR, explicit expressions for the covariant
derivatives of~$K$ and~$M$ in terms of the partial derivatives of
the function~$f$ show that even $\sup\{V_\pm(\cdot,u)|\,u\in I\}$
belongs to~$\si(\Real^2,(1+|x|^\delta)\,dx)$ which gives~\Assiii.
Let us remark that we require the $\si$-integrability in~\AssRiii,
which is a stronger condition than the $\sii$-integrability
required for the second and third derivatives of~$f$. The
assumption~\Assii\/ is a consequence of~\Assiii\/ and the
boundedness of~$V_\pm$.

On the other hand, there is a difference between the
Hamiltonian~(\ref{WCLHamiltonian}) and our operators~$H_\pm$
because the potentials~$V_\pm$ depend themselves on~$\varepsilon$.
Inspecting the proof of Theorem~\ref{Thm.Expansion} we see that
one can modify it to get the expansion~(\ref{expansion})
using $\varepsilon\mapsto V_\pm(\varepsilon)$ as well.
However, expanding the latter in terms of~$\varepsilon$,
the final expansion resulting from~(\ref{expansion}) changes
and to ensure that it represents an eigenvalue
it is necessary that the lowest-order term in this
expansion is not only non-positive but rather strictly
negative. Recall that one can find (physically interesting) examples
of situations when the potential depends on the coupling parameter
in a nonlinear way and a bound state in the critical zero-mean
case may or may not exist~\cite{BCEZ}.

Hence one has to examine carefully the behaviour with respect to
powers of $\varepsilon$ in the expansion obtained
from~(\ref{expansion}) by an appropriate change of the integration
variables,
\begin{eqnarray*}
  w_\pm(\varepsilon)
&=& \frac{\varepsilon}{2\pi} \,\sigma_\pm^2 \int_{\Real^2}
  (V_\pm)_{11}(\sigma_\pm x;\varepsilon) \, dx \\
&+& \left(\frac{\varepsilon}{2\pi}\right)^2  \sigma_\pm^4 \Bigg\{
  \int_{\Real^2\times\Real^2}
  (V_\pm)_{11}(\sigma_\pm x;\varepsilon)
  \left(\Euler+\ln\frac{\sigma_\pm |x-x'|}{2}\right) \\
&&
  \qquad\times\
  (V_\pm)_{11}(\sigma_\pm x';\varepsilon)
  \, dx\,dx'  \\
&-& \sum_{j=2}^\infty \int_{\Real^2\times\Real^2}
  (V_\pm)_{1j}(\sigma_\pm x;\varepsilon)
  \, K_0\left(k_j(\kappa_1)\sigma_\pm |x-x'|\right) \, \\
&& \qquad\times\
  (V_\pm)_{j1}(\sigma_\pm x';\varepsilon) \, dx\,dx'
  \Bigg\}
  +\mathcal{O}(\varepsilon^{2+\gamma}),
\end{eqnarray*}
which determines the possible eigenvalues of~$H_\pm$ via
$E_\pm=\kappa_1^2-\exp(2 w_\pm^{-1})$.
By virtue of~(\ref{km}) we can write
$$
  \varepsilon^{-1}V_2(x,u;\varepsilon)
  =\varepsilon \left(k_0(x)-m_0(x)^2\right)+r_2(x,u;\varepsilon),
$$ where~$r_2$ is an integrable function collecting the terms of
order~$\varepsilon^2$ and higher. Hence
$$
  \forall j\in\Nat: \quad
  \varepsilon^{-1}(V_2)_{1j}(x;\varepsilon)
  =\left[ \varepsilon \left(k_0(x)-m_0(x)^2\right)
  +\mathcal{O}(\varepsilon^2)\right]
  \delta_{1j},
$$
where we abuse the notation a little because the error term
depends on~$x$ as well. The expansion of~$\varepsilon^{-1} v_1$
requires more attention because $\varepsilon^{-1} \Delta_g M$ in
the second term of~(\ref{v1}) is of order one. First of all, we
can write
\begin{eqnarray*}
  \lefteqn{ v_1(x,u;\varepsilon)
  = -u (\Delta_g M)(x;\varepsilon)} \\ &&
  +u^2 \left(\frac{1}{2}\Delta_g K-|\nabla_{\!g}M|_g^2-2M\Delta_g M\right)
  \!(x;\varepsilon)
  +r_1(x,u;\varepsilon),
\end{eqnarray*}
where~$r_1$ is an integrable function of order
$\mathcal{O}(\varepsilon^3)$. This does not mean, of course, that
the first two terms constitute a quadratic polynomial
in~$\varepsilon$, because~$|\cdot|_g$, $\nabla_{\!g}$
and~$\Delta_g$ expand as well. Since the first term in the above
expansion of~$v_1$ is an odd function of~$u$, it does not
contribute to~$(V_\pm)_{11}$. On the other hand, it plays an
important role in the higher modes:
\begin{eqnarray*}
  (v_1)_{11}(x;\varepsilon)
&=& \|u\chi_1\|_{\!\trans}^2
  \left(\frac{1}{2}\Delta_g K-|\nabla_{\!g}M|_g^2
  -2M\Delta_g M\right)\!(x;\varepsilon)
  +\mathcal{O}(\varepsilon^3), \\
  (v_1)_{1j}(x;\varepsilon)
&=& -(\chi_1,u\chi_j)_{\!\trans}\,(\Delta_g M)(x;\varepsilon)
  +\mathcal{O}(\varepsilon^2)
  \quad\textrm{if}\quad j\in\Nat\setminus\{1\}.
\end{eqnarray*}
Notice that the inner product in the second line is non-zero for
even~$j$ only. Now we use the relations
$$
  |\nabla_{\!g}M|_g^2=|\nabla M|^2
  +\varepsilon^2 M_{,\mu}\tilde{\eta}^{\mu\nu}M_{,\nu}
  \qquad\textrm{and}\qquad
  -\Delta_g=-\Delta+\varepsilon^2 L(\varepsilon),
$$
where~$L(\varepsilon)$ is a second-order differential operator
with coefficients which expand as~$\mathcal{O}(1)$.
This together with~(\ref{km}) yields
\begin{eqnarray*}
  \varepsilon^{-1} (v_1)_{11}(x;\varepsilon)
&=& \varepsilon\,\|u\chi_1\|_{\!\trans}^2
  \left(\frac{1}{2}\Delta k_0-|\nabla m_0|^2
  -2m_0\Delta m_0\right)\!(x)
  +\mathcal{O}(\varepsilon^2) \\
  \varepsilon^{-1} (v_1)_{1j}(x;\varepsilon)
&=& -(\chi_1,u\chi_j)_{\!\trans}\,(\Delta m_0)(x)
  +\mathcal{O}(\varepsilon).
\end{eqnarray*}
Since~$C_\pm/C_\mp^2$ and~$\sigma_\pm$
expand like~$1+\mathcal{O}(\varepsilon)$, we arrive at
\begin{eqnarray}\label{WCL.InterExpansion}
  w_\pm(\varepsilon)
&=& \frac{\varepsilon^2}{2\pi} \Bigg\{
  \int_{\Real^2} \left(k_0-m_0^2\right)\!(x)\,dx \nonumber \\
&+& \|u\chi_1\|_{\!\trans}^2 \int_{\Real^2}
  \left(\frac{1}{2}\Delta k_0-|\nabla m_0|-2m_0\Delta m_0\right)\!(x)
  \, dx \nonumber \\
&-& \frac{1}{2\pi} \sum_{j=2}^\infty (\chi_1,u\chi_j)_{\!\trans}^2
  \int_{\Real^2\times\Real^2}
  (\Delta m_0)(x)
  \, K_0\left(k_j(\kappa_1)\sigma_\pm |x-x'|\right) \, \nonumber \\
&& \qquad\times\
  (\Delta m_0)(x')
  \, dx\,dx'
  \Bigg\}
  +\mathcal{O}(\varepsilon^{2+\gamma}),
\end{eqnarray}
where the sum runs in fact over even~$j$ only.

This expression can be further simplified. By a double integration
by parts where one uses the fact that~$f_{,\mu}f_{,\nu\!\rho}\to
0$ as~$|x|\to\infty$ due to~\AssDi\/ and~\AssDii, we get that the
integral of~$k_0$ is equal to zero. We note in this connection
that it follows then from~(\ref{WCLg}) and~(\ref{km}) that the
total Gauss curvature, $\TotK:=\int_{\Sigma_\varepsilon} K
d\Sigma_\varepsilon$, behaves as~$\mathcal{O}(\varepsilon^4)$.
Using the divergence theorem and the assumptions~\AssDi, \AssDii\/
and \AssDiii\/ by which~$\nabla K\to 0$ as~$|x|\to\infty$, we find
that the integral from~$\Delta k_0$ does not contribute as well. A
similar argument employing the Green formula gives
\begin{equation}\label{Green}
  \int_{\Real^2} (m_0\Delta m_0)(x) \, dx
  =-\int_{\Real^2} |\nabla m_0|^2(x) \, dx.
\end{equation}
It is convenient to rewrite the integral part of the last term
in~(\ref{WCL.InterExpansion}) by means of a convolution
\begin{multline*}
\frac{1}{2\pi} \int_{\Real^2\times\Real^2}
  (\Delta m_0)(x)
  \, K_0\left(k_j(\kappa_1)\sigma_\pm |x-x'|\right) \,
  (\Delta m_0)(x')
  \, dx\,dx'  \\
=\left(\Delta m_0,G_k*\Delta m_0\right),
\end{multline*}
where $G_k(\cdot):=(2\pi)^{-1}\,K_0(k|\cdot|)$ and~$k$
abbreviates~$k_j(\kappa_1)\sigma_\pm$. Since~$G_k$ is the
fundamental solution of the distributional equation
$(-\Delta+k^2)G_k=\delta$, we get
\begin{eqnarray}\label{convolution}
\lefteqn{\left(\Delta m_0,G_k*\Delta m_0\right)
  =\left(\Delta m_0,\Delta G_k* m_0\right)
  =\left(\Delta m_0,(k^2 G_k-\delta)*m_0\right)} \nonumber \\
&& =k^2 \left(\Delta m_0,G_k*m_0\right)-(\Delta m_0,m_0)
  =k^2 \left(G_k*\Delta m_0,m_0\right)+\|\nabla m_0\|^2 \nonumber \\
&& =k^2 \left(\Delta G_k*m_0,m_0\right)+\|\nabla m_0\|^2
  =k^2 \left((k^2 G_k-\delta)*m_0,m_0\right)+\|\nabla m_0\|^2 \nonumber \\
&& =k^4 \left(m_0, G_k* m_0\right)
  -k^2 \|m_0\|^2+\|\nabla m_0\|^2.
\end{eqnarray}
In the second line we have employed the identity~(\ref{Green}). If
we insert the obtained expression into the
expansion~(\ref{WCL.InterExpansion}) and use, in addition, the
Parseval identity,
\begin{equation}\label{Parseval}
  \|u\chi_1\|_{\!\trans}^2=\sum_{j=1}^\infty(\chi_1,u\chi_j)_{\!\trans}^2,
\end{equation}
we find that the terms containing~$\|\nabla m_0\|$ cancel, hence
we arrive at
\begin{eqnarray*}
w_\pm(\varepsilon)
  &\!=\!& -\frac{\varepsilon^2}{2\pi} \Bigg\{
  \|m_0\|^2+\sum_{j=2}^\infty (\chi_1,u\chi_j)_{\!\trans}^2\,k^2
  \Big[k^2 \left(m_0, G_k* m_0\right)
  -\|m_0\|^2 \Big]\Bigg\} \\
  && + \mathcal{O}(\varepsilon^{2+\gamma}),
\end{eqnarray*}
where, of course, $k$ expands as~$k_j(\kappa_1)
+\mathcal{O}(\varepsilon)$. Since
$$
  \sum_{j=2}^\infty  (\chi_1,u\chi_j)_{\!\trans}^2\,k_j(\kappa_1)^2
  =\|u\,\partial_3\chi_1\|_{\!\trans}^2-\kappa_1^2\|u\chi_1\|_{\!\trans}^2
  =1,
$$
we see that also the terms containing~$\|m_0\|$ cancel. Using
finally the Fourier transformation as in~(\ref{FourierTrick}), we
obtain
$$
  w_\pm(\varepsilon)=-\varepsilon^2
  \sum_{j=2}^\infty (\chi_1,u\chi_j)_{\!\trans}^2\,k_j(\kappa_1)^4
  \int_{\Real^2} \frac{|\widehat{m_0}(\omega)|^2}
  {|\omega|^2+\kappa_j^2-\kappa_1^2}\,d\omega
  +\mathcal{O}(\varepsilon^{2+\gamma}),
$$
where we have also expanded the remaining~$k$ from~$G_k$
w.r.t.~$\varepsilon$ and included the higher orders into the error
term .

We note that~$m_0$ (and therefore~$\widehat{m_0}$) cannot be
identically zero once~$\Sigma_1$ is not a plane. It can be seen
from the formulae~(\ref{km}): suppose that \mbox{$m_0\equiv 0$}.
Then $f_{,11}\equiv -f_{,22}$, which
yields~$k_0=-({f_{,11}}^2+{f_{,12}}^2)$ and consequently
$k_0\equiv 0$ because we have shown that $\int_{\Real^2}
k_0(x)\,dx=0$. Hence~$K\equiv 0$. At the same time,
$f_{,\mu\nu}\equiv 0$, which gives~$m_1\equiv 0$ and
therefore~$M\equiv 0$. If we thus exclude the trivial planar case,
the lowest-order term in the expansion is strictly negative. Since it is
identical for both~$w_+$ and~$w_-$, it follows
by~(\ref{WCL.Squeeze}) that it is the same also for the expansion
of~$w(\varepsilon)$ in the ground-state energy~$E=\kappa_1^2-e^{2
w^{-1}}$ of the Hamiltonian~$H$. This concludes the proof of
Theorem~\ref{Thm.WCL}.
\begin{rem}
The surface~$\Sigma_1$ is not planar if the
matrix~$\theta_{\mu\nu}$ is not identically zero, \ie, if
~$f_{,\mu\nu}\not\equiv 0$. This can be seen as follows. Suppose
that~$f_{,\mu\nu}= 0$ identically. Solving this as a system of
differential equations, we get~$f(x)=c_\mu x^\mu$, where~$c_\mu$
is a constant vector, which is exactly a plane parametrization.
\end{rem}
%

%%%%%%%%%%%%%%%%%%%%%%%%%%%%%%%%%
\subsection{The Thin Layer Limit}\label{Sec.Thin}
%%%%%%%%%%%%%%%%%%%%%%%%%%%%%%%%%
We say that the layer~$\Omega_\varepsilon$ is thin if
$a\ll\rho_m$. Since the minimum curvature radius introduced in
(\ref{C+-}) explodes as $\varepsilon\to 0$ this condition is
eventually always satisfied. It is useful, however, to consider
also a ``true'' thin-layer limit in which we start with a mildly
curved layer with a fixed $\varepsilon$ which is already in the
asymptotic regime, and make its width $d$ small. To this aim we
rewrite the formula for~$w_1$ in the expansion
$w(\varepsilon)=:\varepsilon^2 w_1
+\mathcal{O}(\varepsilon^{2+\gamma})$ of Theorem~\ref{Thm.WCL} in
a different way. For this we go back to the intermediate
expansion~(\ref{WCL.InterExpansion}). We simplify it as above,
however, do not use~(\ref{convolution}) and~(\ref{Parseval}).
Instead we apply the transformation~(\ref{FourierTrick}) directly
to~$\left(\Delta m_0,G_k*\Delta m_0\right)$, expand~$k$ in terms
of~$\varepsilon$ obtaining thus
$$
  w_1=-\frac{1}{2\pi}\,\|m_0\|^2
  +\frac{\|u\chi_1\|_{\!\trans}^2}{2\pi}\,\|\nabla m_0\|^2
  -\sum_{j=2}^\infty (\chi_1,u\chi_j)_{\!\trans}^2\,
  \int_{\Real^2} \frac{|\widehat{\Delta m_0}(\omega)|^2}
  {|\omega|^2+k_j(\kappa_1)^2}\,d\omega.
$$
An explicit calculation yields
$\|u\chi_1\|_{\!\trans}^2=(\pi^2-6)/(12\kappa_1^2)$, so the second
term in the expression for~$w_1$ is proportional to~$d^2$ by the
definition of~$\kappa_1$. On the other hand, using the estimate
$|\omega|^2+k_j(\kappa_1)^2\geq k_j(\kappa_1)^2
\geq\kappa_2^2-\kappa_1^2=3\kappa_1^2$ together
with~(\ref{Parseval}) and the above explicit result, we see that
the third term is~$\mathcal{O}(d^4)$. Note that this makes sense
because $\|\widehat{\Delta m_0}\|=\|\Delta m_0\|$ is finite due
to~\AssRiii. Summing up the argument, we obtain the
expansion~(\ref{ThinExpansion}).

This formula has a transparent structure.
By virtue of~(\ref{km}),~(\ref{WCLg}),
and the fact that~$\int_{\Real^2} k_0(x)\,dx=0$,
we get the behaviour
$$
  \int_{\Sigma_\varepsilon}\left(K-M^2\right)d\Sigma_\varepsilon
  =-\varepsilon\left(\|m_0\|^2+\mathcal{O}(\varepsilon^2)\right).
$$
Hence the first term of~(\ref{ThinExpansion}) comes from the
surface attractive potential $K-M^2$ which dominates the picture
for thin layers. The last claim follows from~(\ref{v1})
and~(\ref{hatHamiltonian2}), because~$v_1\to 0$ and the leading
term of~$V_2$ is $K-M^2$ in the limit~$u\to 0$,
which one can do as soon as~$d\to 0$. At the same time,
\mbox{$C_\pm\to 1$}, which in view of~(\ref{H1-bound}) leads to the
``limiting'' Hamiltonian $-\Delta_g-\partial_3^2+K-M^2$. Such a
limit is, of course, only formal since the transverse spectrum
explodes, which classically corresponds to increasingly rapid
oscillations in the transverse direction. Nevertheless, a similar
limit with the confinement realized by a harmonic potential
transverse to a compact curved surface was discussed recently in a
rigorous way in~\cite{FrHe} and led to the same tangential
operator~$-\Delta_g+K-M^2$.

%%%%%%%%%%%%%%%%%%%%%%%%%%%%%
\subsection*{Acknowledgments}
%%%%%%%%%%%%%%%%%%%%%%%%%%%%%
The authors wish to thank to Pierre Duclos for helpful discussions
and for the hospitality in Centre de Physique Th\'eorique,
Marseille-Luminy, where a part of the work was performed. The
research has been partially supported by the Grant GAAS A1048101.
%
%\newpage
%\bibliography{mycites}
%\bibliographystyle{amsplain}
%
\providecommand{\bysame}{\leavevmode\hbox to3em{\hrulefill}\thinspace}

\end{document}